%% file: main.tex
\documentclass[sigconf]{acmart}

\usepackage{graphicx}
\usepackage{colortbl}
\usepackage[normalem]{ulem}
\useunder{\uline}{\ul}{}
\usepackage{array}
\usepackage{tabularx}
\usepackage{booktabs}
\usepackage{lscape}
\usepackage{multirow, makecell}

\AtBeginDocument{%
  }

\copyrightyear{2026}
\acmYear{2026}
\setcopyright{cc}
\setcctype{by}
\acmDOI{10.1145/3800645.3813053}

\acmConference[DIS ’26]{Designing Interactive Systems Conference}{June 13--17, 2026}{Singapore, Singapore}
\acmBooktitle{Designing Interactive Systems Conference (DIS ’26), June 13--17, 2026, Singapore, Singapore}
\acmISBN{979-8-4007-2563-0/2026/06}

\begin{document}

\title{How Designers Envision Value-Oriented AI Design Concepts with Generative AI}


\author{Pitch Sinlapanuntakul}
\orcid{0000-0003-3551-8531}
\affiliation{%
  \department{Human Centered Design \& Engineering}
  \institution{University of Washington}
  \city{Seattle}
  \state{WA}
  \country{USA}
}
\email{wspitch@uw.edu}

\author{Aayushi Dangol}
\orcid{0009-0000-0837-9738}
\affiliation{%
  \department{Human Centered Design \& Engineering}
  \institution{University of Washington}
  \city{Seattle}
  \state{WA}
  \country{USA}
}
\email{adango@uw.edu}

\author{Xiaoyi Xue}
\orcid{0009-0002-1930-471X}
\affiliation{%
  \department{Human Centered Design \& Engineering}
  \institution{University of Washington}
  \city{Seattle}
  \state{WA}
  \country{USA}
}
\email{xxy0119@uw.edu}

\author{Mark Zachry}
\orcid{0000-0002-1067-7168}
\affiliation{%
  \department{Human Centered Design \& Engineering}
  \institution{University of Washington}
  \city{Seattle}
  \state{WA}
  \country{USA}
}
\email{zachry@uw.edu}

\renewcommand{\shortauthors}{Sinlapanuntakul et al.}

\begin{abstract}
As AI integrates into design practice, designers increasingly use generative AI tools to envision AI-enabled solutions, positioning AI as both design tool and design material. This dual role creates recursive value tensions distinct from traditional design work. We engaged 18 designers in a concept envisioning activity and interviews to understand how they navigate values and recognize potential harms in this context. Our analysis reveals that (i) designers engage in reciprocal reflection-in-action with AI; (ii) this process surfaces multi-level value tensions across tool, designer, and concept; (iii) designers demonstrate greater attunement to harm recognition as a primary design signal than to articulating positive value fulfillment; and (iv) designers exercise anticipatory judgment through meta-design reasoning about how tool assumptions risk propagating into designed concepts and future use contexts. We extend Schön’s reflection-in-action framework and discuss implications for redesigning AI-mediated design tools, supporting harm-centered reasoning, and positioning design as foundational to AI development.
\end{abstract}




\maketitle

\section{Introduction}

Designers increasingly use generative AI tools to conceptualize and shape ideas for AI-enabled products and services. These tools promise speed, creative inspiration, and access to technical possibilities that might otherwise remain opaque. Yet they introduce a fundamental challenge. When designers use AI to envision AI-enabled solutions, they must simultaneously navigate value considerations while remaining attentive to potential harms manifested in the AI-generated suggestions and concept solutions \cite{namer2025harm96designers}. Early ideation decisions carry particular weight because the framings, assumptions, and value commitments established at this stage tend to persist through later development phases. The process of AI concept envisioning is crucial in shaping both the functional capabilities and fundamental identity of AI-enabled solutions \cite{yang2018UXML, yildirim2022, zdanowska2022, zimmerman2020}. During this formative stage, designers establish not merely what these AI-enabled solutions can do, but their purpose, scope, and relationship to human flourishing \cite{nakata2025}. Early framing decisions about design scope, including which values are centered, have shaped documented harms in AI-enabled products, from biased hiring screening to opaque content moderation \cite{raji2020}. This illustrates why the approach designers take to value integration at this formative stage carries critical downstream implications.

The challenge is distinctive because values operate at multiple levels simultaneously. Design practice is inherently anchored in human values that manifest throughout the design process and become embedded in technological outcomes \cite{friedman2019, nelson2014}. Values are not only what designers aim to embed in systems; they also reside within the tools designers use. When a designer employs a generative AI tool, that tool carries embedded assumptions, training data distributions, and architectural constraints. In this context, those assumptions do not merely affect the designer’s reflective process but also risk propagating into the AI-enabled concept being designed. When a generative AI tool’s defaults favor certain values like efficiency, particular interaction paradigms and orientations become candidate design ideas or unexamined omissions in the AI-enabled solution being envisioned. This propagation pathway distinguishes the present context from using AI to support conventional design tasks. The normative commitments embedded in the ideation tool may resurface as normative commitments in the resulting AI that end users eventually interact with, in ways that are difficult to trace back to early ideation decisions. In early-stage ideation, three value streams intersect and sometimes collide, including the designer’s stated intentions, the AI tool’s interpretations and affordances, and the values imagined for the resulting solution. This convergence creates novel tensions that differ fundamentally from traditional design contexts, where materials respond predictably to manipulation and do not actively interpret designer intent.

Prior research illuminates aspects of this challenge but leaves significant gaps. Studies of generative AI in design have examined these tools as creative assistants, focusing on productivity gains, inspiration, or collaborative potential. Separately, a rich literature on value-oriented and ethical design has articulated principles, frameworks, and methods for embedding values into technological systems \cite{friedman2019}. Design practitioners who specialize in crafting the interactions between users and digital products possess sophisticated abilities in identifying user needs and shaping meaningful interactions. However, they often lack specialized AI expertise, making it difficult to translate abstract AI affordances into concrete, value-aligned concepts \cite{subramonyam2022, yang2018UXML, yildirim2023pairguidebook}. Current design frameworks and resources predominantly focus on implementation excellence rather than supporting the foundational work of framing appropriate problems and centering solutions around human values \cite{googlepair, microsoftguidelines, ibmguidelines, appleguidelines, yang2019, yildirim2023pairguidebook}. What remains empirically underexplored is what happens when these concerns converge in actual practice, particularly when designers use generative AI tools to conceptualize AI systems while actively attempting to center human values and recognize potential harms during the earliest conceptual phases. Additionally, much foundational design theory, including seminal accounts of reflection-in-action, assumes materials that respond transparently and predictably to designer manipulation. Generative AI disrupts this assumption through interpretive agency and inherent opacity \cite{subramonyam2022, yang2018UXML}.

To investigate these dynamics, we conducted an exploratory study with 18 designers spanning UX/product design, service design, and design research, all with prior experience integrating AI into their design work. Using a concept envisioning activity with generative AI and reflective interviews, we examined how designers navigated value considerations and recognized potential harms during early conceptualization of AI-enabled solutions. By \textit{value-oriented}, we mean that participants were asked to anchor their design activity in a single self-selected human value and intentionally center it throughout ideation. This is distinct from conducting a formal Value Sensitive Design (VSD) process; the activity was not structured to support iterative investigation of relevant parties, nor did it require participants to systematically map or negotiate competing values across direct and indirect parties as VSD’s stakeholder analysis prescribes \cite{friedman2002, friedman2019}. Participants were not required to have prior VSD experience, and the single-session format does not approximate the depth of continuous value engagement VSD entails. Rather, the activity draws conceptual inspiration from VSD’s commitment to proactive value integration while examining how designers naturally engage with value considerations within the constraints of a generative AI-mediated ideation context. Because existing research has not examined how the dual role of AI, as both design tool and design material, creates value negotiation processes that differ from traditional practice, we address the following research questions:

\begin{enumerate}
    \item[]\textit{RQ1}: How do designers engage with human values and generative AI tools in early-stage AI concept envisioning?
    \item[]\textit{RQ2}: How does this engagement, where AI functions as both tool and perceived design material, influence designers’ value articulation and recognition of potential harms?
\end{enumerate}

Our study reveals several key findings. First, designers engaged in reciprocal reflection-in-action, a bidirectional dialogue where AI functioned as a responsive material that actively interpreted designer intent rather than merely responding to manipulation, extending Schön’s framework \cite{schon1987}. Through this engagement, designers surfaced multi-level value tensions spanning tool, designer, and concept. Critically, participants demonstrated greater attunement to recognizing potential harms and misalignments than to articulating positive value ideals. Our findings further surface participants’ speculation that design expertise could be reoriented from execution-focused practice toward interpretation-centered engagement. As AI reshapes design practice, understanding how designers navigate this recursive value-negotiation process becomes crucial for supporting responsible innovation. Our study contributes to this understanding by examining the novel condition in which AI functions both as a design medium (\textit{designing with AI}) and as the object of design inquiry (\textit{designing for AI}), revealing critical interdependencies between AI-mediated practice and concept development.


\section{Background and Related Work}

Drawing from studies in human-computer interaction, design studies, and AI ethics, this section outlines challenges in AI concept design, the material properties of AI in design practice, and approaches to value considerations in AI development.


\subsection{Challenges of Designing AI Concepts}

Design practitioners have long occupied a pivotal role in problem framing and innovation conceptualization. Yet their involvement in AI development frequently occurs after fundamental decisions about system purpose and scope have been established \cite{dove2017, jung2025, yildirim2024}, significantly constraining their capacity to influence what should be designed rather than merely how to design it effectively \cite{saxena2025, yildirim2022}. Research demonstrates that when designers are excluded from early-stage concept envisioning, resulting AI solutions may achieve technical sophistication while failing to address appropriate problems or align with human values \cite{dove2017, nishal2025}.

Designers are often excluded from early problem-framing stages where values are established \cite{benjamin2021, feng2023, saxena2025, yang2020, yang2018UXML}. This exclusion compounds the challenges of AI concept envisioning, particularly for designers who lack specialized AI expertise. Designers must translate abstract AI affordances, which embody probabilistic behavior and complex sociotechnical implications, into meaningful solutions that address authentic human needs \cite{benjamin2021, dove2017, park2025, subramonyam2022, wiener1988, yang2019, yildirim2023pairguidebook}. Unlike conventional design materials with well-defined constraints and affordances, AI systems are difficult to conceptualize, communicate, or anticipate in early ideation. Empirical research demonstrates that designers frequently develop inaccurate mental models of AI capabilities, often envisioning concepts that exceed technical feasibility or failing to account for emergent effects and value tensions \cite{nishal2025, yang2018UXML, yang2020, yildirim2024}.

Current design frameworks and resources predominantly focus on what Buxton \cite{buxton2010} characterizes as \textit{designing things right} rather than \textit{designing the right thing}. Existing resources from major technology companies \cite{googlepair, microsoftguidelines, ibmguidelines, appleguidelines} emphasize technical refinement and user experience optimization while providing limited guidance for the foundational work of problem identification and value alignment during concept development \cite{yang2019, yildirim2023pairguidebook}. While some frameworks attempt to translate technical AI functionality into design-appropriate conceptual models \cite{jansen2023, yildirim2022, yildirim2023designresources, yildirim2023pairguidebook}, and others provide concrete examples of AI applications to scaffold understanding \cite{park2025, yildirim2023designresources}, product teams primarily use these resources for refinement and implementation rather than early-stage ideation \cite{yang2018UXML, yang2018mappingML, yildirim2022, yildirim2023pairguidebook}. This gap between available resources and actual design needs highlights the continued challenge of supporting value-oriented AI concept development, particularly as designers increasingly work with generative AI tools during conceptualization.


\subsection{AI as Design Material}

Understanding AI’s distinctive properties as a design material is essential for grasping how concept development differs when designers work with AI systems \cite{holmquist2017, lindgren2023}. A defining characteristic of AI as a design material is its dual nature, functioning simultaneously as both a tool within design processes and as the material of design itself \cite{dove2017, yang2020}. This duality creates recursive relationships between design decisions and material properties, where AI systems evolve through interaction and deployment in ways that transcend initial design specifications \cite{holmquist2017}. Consequently, designers must conceptualize dynamic systems capable of adaptation and emergent behavior rather than static artifacts \cite{yang2020, zimmerman2020}.

AI materiality introduces distinctive forms of uncertainty into the design process. Unlike materials with stable, predictable properties, AI systems operate probabilistically and may produce unexpected outputs or behaviors that diverge from designers’ intentions \cite{benjamin2021, yang2018UXML, yang2020, yildirim2024}. This uncertainty complicates conceptualization by requiring designers to anticipate broader ranges of potential system behaviors and consequences \cite{benjamin2021, feng2023}. The probabilistic nature of AI challenges design practices that assume deterministic relationships between inputs and outputs \cite{holmquist2017, yang2018UXML}.

Bidirectional, improvisational engagement between practitioner and material is not unique to AI. Schön’s concept of \textit{backtalk} describes how conventional materials respond to a designer’s moves in ways that surprise, reveal constraints, and prompt reframing \cite{schon1983, schon1992}. Ingold similarly argues that skilled craft practice involves following and responding to materials as active participants rather than imposing predetermined form upon passive matter \cite{ingold2007}. Expert craftspeople already engage with materials in reciprocal, interpretively rich ways. What distinguishes AI as a design material is therefore not the presence of bidirectionality per se, but the specific character of that responsiveness. Rather than returning force through physical resistance and structural affordances that characterize backtalk in craft practice, AI systems return semantic interpretation and generative agency \cite{holmquist2017, lindgren2023}. They infer designer intent, propose novel directions, and embed normative assumptions into outputs in ways that place evaluative and ethical demands on designers that Schön’s framework, developed in the context of physical and diagrammatic materials, did not anticipate.

The material properties of AI also raise fundamental questions about agency and control. As AI systems gain capabilities for autonomous decision-making and adaptation, designers must reconsider their role in shaping system behavior \cite{buchanan1992, giaccardi2015, lindgren2023}. This shift from designing fixed artifacts to designing adaptive systems requires conceptual frameworks that account for distributed agency among human designers, AI systems, and users \cite{giaccardi2015, lindgren2023, verbeek2011}. The temporal dimension of AI materiality further complicates concept design as AI systems evolve over time through data accumulation and learning, creating dynamic relationships between initial design decisions and eventual system behavior \cite{giaccardi2015, holmquist2017}. This temporal extension challenges designers to conceptualize not just immediate functionality but long-term trajectories of system evolution \cite{holmquist2017, lindgren2023, yang2020}.

Recent studies of how designers work with generative AI tools reveal emerging forms of human-AI collaboration in creative processes where AI systems act as creative partners rather than passive tools \cite{palani2024, shi2023, takaffoli2024, verganti2020, xu2019}. These collaborations raise questions about authorship, creativity, and the distribution of agency in design processes. However, existing work has not examined how these collaborative dynamics intersect with value considerations when the AI tool being used and the AI-enabled solutions being designed are of the same type.


\subsection{Values as Enacted in Early-Stage Design Practice}

Values in design cannot be treated as abstract principles appended at the conclusion of a process; they are enacted through the ways designers frame problems, negotiate constraints, and evaluate what counts as a \textit{good} course of action. Work in VSD demonstrates that integrating values from the outset produces outcomes qualitatively different from those that treat values as compliance checks or post-hoc requirements \cite{friedman2002, friedman2019, umbrello2021}. Yet much of this literature concentrates on identifying and categorizing the values of relevant parties, while leaving underexplored how values operate in the uncertain and iterative moments of early-stage design, precisely when assumptions about users, systems, and societal consequences first take shape \cite{friedman2019, nelson2014}. Values are further situated in context and temporality \cite{verbeek2011, dignum2019, raji2019}. Decisions made during early conceptualization, whether they privilege efficiency over reflection, personalization over privacy, or automation over human agency, cascade through technical implementations and social practices, shaping system behaviors and broader societal outcomes. The adaptive and emergent character of AI amplifies these effects, demanding foresight not only into immediate user interactions but also into how value commitments may shift as systems scale and intersect with complex socio-technical environments \cite{gabriel2020, shen2024bidirectional, shen2024valuecompass, sadek2024guidelines, stray2024}.

The introduction of AI into the design process complicates value enactment. AI tools carry their own embedded value orientations and technical authority, which interact recursively with designers’ guiding intentions and the values intended for the concept itself \cite{dignum2019, raji2020, ryan2020, shi2023, sinlapanuntakul2025SLR}. In early-stage ideation, AI’s suggestions shape the conceptual space while designers’ interpretations and selective incorporation of outputs simultaneously influence which aspects of the tool’s embedded assumptions are realized. This dynamic generates novel tensions and opportunities, which makes friction and emergent harm a critical locus for value-sensitive reasoning, even when explicit articulation of ethical ideals is absent \cite{gabriel2020, whittlestone2019}.

Despite the proliferation of industry guidelines emphasizing transparency, trust, and usability \cite{googlepair, microsoftguidelines, ibmguidelines, appleguidelines}, support for early-stage, value-oriented concept development remains limited. Emerging research has begun to address this gap by proposing frameworks that help designers orient AI concepts around specific values while anticipating broader societal impacts \cite{shen2024bidirectional, sinlapanuntakul2026toolkit, sadek2024guidelines}. These approaches foreground AI not merely as a production tool but as interpretive material capable of surfacing assumptions, revealing latent tensions, and enabling negotiation of value trade-offs. Conceptualizing values as generative, contextually enacted, and recursively negotiated rather than fixed principles provides a foundation for understanding how design practice mediates multi-level interactions. It positions early-stage design as a critical moment where value commitments become materially and socially instantiated \cite{friedman2019, nelson2014, umbrello2021}.

While VSD offers a rigorous tripartite methodology encompassing conceptual, empirical, and technical investigations across parties and system states \cite{friedman2002, friedman2019}, our study does not operationalize this apparatus. We also do not assume design is inherently value-driven. Explicit value reflection is far from universal in professional practice, and VSD remains a contested and unevenly adopted framework \cite{borning2012, shilton2018}. Our study draws on VSD’s foundational orientation toward proactive value integration as a conceptual lens, treating value engagement as naturally occurring practice to be examined rather than as structured VSD activity.


\section{Method}

In this section, we describe our conceptual analysis of relevant parties \cite{friedman2013} and our conceptual design activity. We follow this with information about our semi-structured interviews, including participants, procedure, and analysis.


\subsection{Analysis of Relevant Parties}

To orient our investigation of value-related implications, we conducted a conceptual analysis of relevant parties (see \hyperref[tab:relevant-parties-analysis]{Table~\ref*{tab:relevant-parties-analysis}} in \hyperref[appendix:relevant-parties-analysis]{Appendix~\ref*{appendix:relevant-parties-analysis}}), drawing from selected value-sensitive design literature (e.g., \cite{li2024,popa2021,shi2023, knearem2023}). We identified two groups: direct parties (i.e., those who interact directly with AI design tools) and indirect parties (i.e., those who may be impacted downstream by AI-influenced design decisions). For each group, we outlined associated values and applied the VSD approach to conceptually examine potential value tensions that might arise as AI use in design disrupts processes and practices. This analysis served not as empirical data but as scaffolding for understanding which values to attend to in interviews. This analysis yielded two primary outcomes. First, it informed an initial construction of the value list presented to pilot participants in the design activity (see \hyperref[design activity]{Section~\ref*{design activity}}) by identifying values relevant across both direct and indirect parties. Second, it shaped the interview guide by identifying value tensions that served as probes for eliciting reflection beyond the specific activity.


\subsection{Design Activity} \label{design activity}

We developed a future-oriented design scenario to prompt participants’ iterative engagement with generative AI under loosely constrained, value-focused conditions. The prompt underwent iterative refinement based on feedback from three experienced design experts. We then piloted the prompt in a design jam with 11 experienced designers \cite{sinlapanuntakul2025designjam}. We observed that a 15-minute window was effective for ideation and rough concept sketching, leading us to allocate 20 minutes for the final activity to allow adequate time for interacting with AI, concept ideation, and sketch creation. This timing balanced exploratory freedom with focused conceptual development while preserving the integrity of each participant’s reflection. The finalized prompt encouraged participants to go beyond conventional design constraints: 

\begin{quote}
    “Imagine a future, 3 years from now, where AI plays a significant role in design. You are a product designer tasked with designing a novel AI-powered technology that supports designers in their work. This technology can be anything you envision—a physical product, a digital interface, or a combination of both—and it doesn’t have to be web- or mobile-based. Consider both the capabilities and limitations of this technology as you design it. You are allowed to break free from preconceived notions and redefine the potential of AI for design.”
\end{quote}

The activity was intentionally future-oriented to surface participants’ reflective reasoning processes when working under unfamiliar constraints, leveraging their experiential knowledge and reflective practice as a form of situated cognition to articulate technological possibilities through established professional practice. By asking designers to envision AI-enabled concepts 3 years in the future, we encouraged innovative thinking while maintaining connection to near-term technological advancement. We also structured the activity for individual completion to isolate personal reflection processes, independent of group dynamics or project-specific organizational goals. While this differed from the collaborative nature of some design practices, it allowed designers to think with and about generative AI, negotiating between leveraging AI capabilities and envisioning appropriate interaction modalities guided by value commitments rather than external constraints. 

To ground concepts in human values, participants selected one value from a predefined list (i.e., accountability, autonomy, creativity, efficiency, growth, identity, ownership, privacy, sustainability, and trust) drawn from Friedman et al. \cite{friedman2013} and our analysis of relevant parties, and integrated it as a core design commitment. This list, modified through pilot testing with four design practitioners and reflective discussions among two authors to ensure relevance to design-led contexts and a more holistic envisioning of AI concepts, included values critical for ethical and societal considerations as well as individual and collective flourishing. 

We invited participants to self-select their value rather than assigning one, to preserve the authenticity of their priorities and choices, mirroring how designers naturally orient toward values in practice \cite{friedman2019,nelson2014} and providing ecological validity to our findings. We constrained participants to a single value, rather than asking them to navigate multiple competing values simultaneously, to maintain tractability within the 20-minute session and to allow observation of how a single value commitment evolves, gets reinterpreted, and potentially generates secondary tensions through AI-mediated ideation. We acknowledge this departs from the multi-value navigation VSD recommends and discuss potential implications in \hyperref[limitations]{Section~\ref*{limitations}}. While we presented values as fixed guiding labels for clarity, they were not static in our activity. They became more specific, nuanced, and sometimes reinterpreted through ideation, sketching, and interactions with AI tools.


\subsection{Participants}

We recruited 18 interaction designers through diverse professional networks, including X (formerly Twitter), LinkedIn, and HCI- and design-specific communities on Slack and Discord. Our recruitment targeted design practitioners working in UX design, product design, service design, and design research, with at least two years of professional experience and prior use of generative AI in design practice. Participants ranged in age from 22 to 51 years (\textit{M} = 32.11, \textit{SD} = 8.47), with 2 to 24 years (\textit{M} = 7.42, \textit{SD} = 7.14) of professional experience. Four had direct professional experience designing AI-enabled products, while the remaining 14 had experience designing AI through personal projects or in academic settings. This composition ensured all participants had relevant expertise for conceptualizing AI-enabled design tools. Detailed participant information is provided in \hyperref[tab:participants]{Table~\ref*{tab:participants}}.

\input{materials/participants}


\subsection{Procedure}

We conducted 70-minute virtual sessions via video conferencing. This study was approved by our University’s Institutional Review Board (IRB). Participants received a \$35 gift card as compensation. Each session comprised two sequential parts: (1) a value-oriented design activity and (2) a semi-structured interview.

In part 1, participants received the design prompt on-screen and selected one core value from the provided list. They were instructed to use familiar generative AI tools (e.g., ChatGPT, Claude, Copilot, Gemini, Midjourney, Perplexity) for ideation support but not to paste the design prompt directly into AI systems. While conceptually informed by VSD’s argument for anchoring design in human values from the outset \cite{friedman2019}, the activity does not constitute a VSD process. Participants engaged with a single self-selected value within a bounded single-session format, without the iterative investigation of relevant parties or recursive value tension mapping VSD entails, and none were required to have prior VSD experience. Our procedure is thus better understood as a value-oriented design probe structured to surface how designers naturally engage with value considerations under generative AI assistance. During the 20-minute activity, participants created concept sketches capturing their design’s purpose, features, and intended value integration while engaging in think-aloud protocol throughout, with researchers taking notes to document emerging reasoning and decision-making. Following completion, they photographed or screenshotted their sketches for submission, which we later re-sketched for clarity (\hyperref[fig:concept-sketches]{Figure~\ref*{fig:concept-sketches}}).

\input{materials/concept-sketches}

In part 2, we conducted 45-minute semi-structured interviews, audio-recorded with consent. Interviews began with participants’ backgrounds and AI-in-design experiences, establishing a foundation for broader discussion beyond the specific activity. We then asked questions about their specific design solutions, AI tool integration processes, value manifestation in their broader design work, and speculations about AI’s future role in design practice. Rather than focusing exclusively on the completed activity, we used it as a probe to elicit reflection on participants’ professional experiences, challenges, and aspirations beyond the immediate task. Following the interview, participants were debriefed, thanked, and compensated for their time and participation.


\subsection{Data Analysis}

After completing all the interviews, we used grounded theory \cite{groundedtheory-glaser, groundedtheory-muller} to analyze the perceptions and design practices of participants in relation to AI. Two authors independently open-coded two interview transcripts to identify preliminary concepts and emerging patterns. Based on this preliminary coding, the first author developed an initial codebook, which captured key codes and definitions grounded in participants’ reflections and experiences. Next, two authors applied the initial codebook to code two additional interview transcripts, using a segment-based approach to test the robustness and clarity of the codes. Following this round of coding, the two authors discussed and refined the codebook through constant comparison, addressing any ambiguities or overlaps in the codes. This process led to a final version of the codebook (see \hyperref[tab:codebook]{Table~\ref*{tab:codebook}} in \hyperref[appendix:codebook]{Appendix~\ref*{appendix:codebook}}), which more precisely captured the range of information in participants’ responses.

Once the final codebook was established, four authors used it to code all 18 interview transcripts in a two-stage process. One author initially coded each transcript segment by segment, followed by a review from a second author who flagged and re-coded segments where interpretations diverged. In pairs, we discussed and reconciled any discrepancies in our coding to reach a shared consensus and understanding. We then collaboratively developed the coded data into conceptual categories and relationships that reflected insights from the interviews, capturing the underlying values, motivations, and perspectives that shaped participants’ practices and perceptions of AI and design. This interpretive approach acknowledges the role of researchers’ positionality in developing conceptual frameworks, and our diverse research team brings expertise across UX of AI, child-AI interaction, creativity support tools, HCI theory, and communicative practices, all grounded in human-centered design approaches. 


\section{Findings}

In this section, we present findings organized around three areas that emerged when designers used generative AI to develop AI-enabled concepts within a single-value constraint. We document not only the difficulties but the strategies participants improvised to navigate them, and note where those strategies were productive and where they remained inconsistent or insufficient. We present these patterns as empirical descriptions of how designers navigated value considerations under a specific set of constraints (i.e., single-value framing, limited session duration, and the AI-generated suggestions) but not as evidence of successful VSD practice or as reliable models for value-aligned AI concept development.


\subsection{Challenges and Strategies in Engaging with AI as Interpretive Material}

To ground the findings that follow, we briefly characterize two prototypical patterns of engagement. The first, most frequent among participants who selected efficiency, involved relatively rapid convergence. They articulated their value, engaged generative AI for ideation, received technically coherent suggestions that appeared value-aligned, and moved into concept sketching with limited friction. Reflection on what this process may have foreclosed emerged primarily in the subsequent interview. The second, observed most consistently among participants who selected values in greater tension with AI tool defaults (e.g., creativity, sustainability, autonomy, and growth), involved more sustained difficulty, including repeated prompt reformulation and explicit rejection of AI suggestions. These participants described greater difficulty completing the \textit{right} sketches but also more iterative engagement with their value commitments, and were more likely to articulate, unprompted, what the AI could not do, naming specific qualities of their value that resisted translation into generative outputs. These are not discrete types but poles of a spectrum that provide context for our findings.

\subsubsection{Value Selection and Translation as a Situated Practice}

Designers approached AI concept development through value selections rooted in lived experiences rather than abstract frameworks. All 18 participants contributed sketches and selected foundational values based on situated frustrations or aspirations from their professional contexts, with 15 concept sketches presented in \hyperref[fig:concept-sketches]{Figure~\ref*{fig:concept-sketches}} (three were excluded due to illegibility of the original sketches). While participants from smaller studios and organizations (e.g., P13, P15, P16) prioritized \textit{creativity} and \textit{sustainability} and academically oriented participants (e.g., P18) selected \textit{growth}, the predominant choice across the group was \textit{efficiency}, suggesting that designers’ imaginative range is anchored by current AI framings.

Efficiency is the default orientation already embedded in most digital product frameworks and in the generative AI tools participants used for ideation. That a majority gravitated toward it even when given free choice from a list of 10 values suggests their imaginative space had been shaped by tools they routinely work with. Efficiency is structurally compatible with AI’s existing generative capabilities, produces little friction with AI tool defaults, and tends to extend existing technological trajectories rather than challenge them. Participants who selected it entered ideation with a value the AI could fluently service, with less friction available to interrupt early convergence. The efficiency concentration therefore reflects not only individual professional orientation but a structuring constraint embedded in AI tool framings themselves, which appear to limit which values feel available, communicable, and actionable during ideation. Participants who selected non-efficiency values frequently described needing to reframe their value before the AI could engage with it at all, a translational effort that itself shaped and in some cases constrained the concepts that ultimately emerged.

P11’s selection of \textit{efficiency} exemplifies this embodied approach. They explained, “\textit{I sketched the problems I faced at my studio, like fast-paced deliverables and poor resource allocation. So, when I saw [efficiency], I immediately picked that because it allows me to operationalize other values like trust and creativity in both the AI I’m using and the AI I’m designing}.” Values thus function as hierarchical systems where foundational choices enable secondary value manifestations. Similarly, P18 connected \textit{growth} to personal learning aspirations while acknowledging professional pressures, describing that “\textit{I’m a lifelong learner. In terms of my profession as a designer, I personally don’t just want to be stagnant, even with job security. But I also see how the industry is pushing toward AI integration, so growth feels like survival too}.” This dual motivation of intrinsic curiosity coupled with professional adaptation characterized value selection as authentic expression rather than explicit functional criteria. In both cases, the selected value carried prior situational meaning, functioning less as a neutral starting point and more as an already-interpreted commitment that shaped how AI outputs were subsequently read and accepted or rejected. Across participants, value selection operated as an ontological commitment, establishing the lens through which AI possibilities are explored.

\subsubsection{Reciprocal Reflection-in-Action}

When designers work with conventional materials, they engage in reflection-in-action, observing consequences of actions and adjusting approaches based on feedback. With generative AI, a fundamentally different dynamic emerged. Unlike conventional materials like paper, wood, or clay that respond mechanistically to designer manipulation, AI actively interpreted designer intent and generated novel suggestions, creating a feedback loop that was bidirectional and interpretive rather than unidirectional and mechanical. We observed a recurring pattern of reciprocal reflection-in-action that unfolded across three interconnected modes (see \hyperref[fig:RRiA]{Figure~\ref*{fig:RRiA}}). These modes are not strictly sequential but flexible, and designers moved fluidly between them based on conceptual needs. 

\input{materials/RRiA}

P4 further captured this complexity, explaining “\textit{When using AI to design AI, I’m constantly toggling between critiquing how well the AI understands AI capabilities and using those outputs to shape my own AI concept. For me, this isn’t present when using other non-AI tools}.” P11 described the multi-layered nature, noting “\textit{I’m watching how ChatGPT interprets my efficiency value, then how it suggests efficiency features for my AI concept, then I’m judging whether those suggestions actually align with what I intend by efficiency. Each step teaches me something about the others}.” This multi-layered reflection created technological self-reflection where the tool became both medium and material of design inquiry. Rather than proceeding through linear action, observation, and adjustment, designers engaged in recursive cycles where interpretation at one level continuously informed judgment at another.

\paragraph{Value externalization} Designers translated embodied and intuitive understanding of their selected value into explicit and communicable language suitable for AI engagement. This translation was non-trivial. Sixteen of 18 participants struggled to translate abstract value commitments into concrete and communicable ideas. As P13 put it, “\textit{I know I want my AI solution to boost users’ creativity, but what does that even mean in terms of interactions and interfaces? I kept relying on abstract words that were hard to turn into actual design features}.” The externalization process forced designers to surface and clarify ambiguities in their own thinking. For example, P15, whose background is in child-centered design, described this dynamic during the think-aloud phase while reformulating prompts around creativity multiple times, stating “\textit{I started with creativity as my core value, but after I tried to explain my ideas to ChatGPT repeatedly, I realized I actually meant something more specific. It’s like the ability to make unexpected connections between distant concepts. The translation process helped clarify my own thinking}.” Think-aloud data corroborated this pattern. The act of formulating and articulating value commitments into prompts for AI demanded a level of specificity and operationalization that prior reflection alone had not produced. This extends accounts of external representations as thinking instruments. Here, the representation is a communicative act addressed to a system that would interpret and act on it, introducing a productive constraint different from sketching or note-taking.

\paragraph{Interpretive encounter} Designers engaged with AI-generated outputs as provocative interpretations that revealed both alignments and gaps relative to their intentions. Critically, designers did not treat AI outputs as finished work but as suggestions that invite critical evaluation, a tension that surfaced repeatedly across participants. P3 described this tension, stating “\textit{the AI generated ideas about AI functionalities seemed both technically plausible and aligned with the value I chose. But I later realized that it created a false sense that I had thoroughly explored the solution space when I, in fact, hadn’t}.” This revealed an important tension that designers knew they should engage critically with AI suggestions, yet the combination of technical sophistication and apparent value alignment created pressure toward acceptance. Examination of the concept sketches in \hyperref[fig:concept-sketches]{Figure~\ref*{fig:concept-sketches}} further deepens this observation. A substantial proportion of efficiency-labeled sketches share structural features, like workflow automation interfaces and AI-assisted task prioritization that closely resemble the genres described in major technology company design guidelines \cite{microsoftguidelines, appleguidelines, ibmguidelines, googlepair} and, by extension, the outputs generative AI most readily produces for efficiency-framed prompts. This convergence in concept structure suggests the interpretive encounter produced concept directions shaped more substantially by AI outputs than participants recognized during the activity. Surface value alignment (i.e., AI suggesting efficiency-oriented features for an efficiency-value prompt) may have masked a narrowing of the design space rather than an exploration of it. P7 also described the generative potential, noting “\textit{The act of using AI to help think about AI forced me to articulate not just what I valued, but why I valued it and how it should be designed differently across all my ideas}.” While this is true for many, a couple of participants found this pressure to explicitly formulate value reasoning as “\textit{cognitively taxing}” (P9 and P11) rather than clarifying, particularly when iterative prompt refinement failed to converge on outputs that felt meaningfully aligned with their intentions.

\paragraph{Reflective redirection} Designers critically evaluated AI suggestions and used both alignment and misalignment as stimuli for deeper thinking. During this mode, designers made deliberate decisions about whether to accept, modify, or reject contributions based on value alignment and conceptual coherence rather than technical sophistication alone. As P14 described, “\textit{I tried to jump from explaining my selected value and initial idea to getting AI concept suggestions, and I grabbed the first coherent response. But then I forced myself to sit with the suggestions and asked some questions. That was when I found problems I initially missed.}” Deliberately introducing pauses emerged as a strategy that helped designers maintain critical perspective. Similarly, P12 “\textit{realized that ChatGPT’s suggestions reinforced existing AI interaction patterns. Breaking free of this loop required deliberate effort to imagine fundamentally different AI paradigms}.” This capacity for distancing was visible in sketch-level differences as well. Participants who described forcing pauses before accepting AI suggestions produced sketches with more explicit notations of constraints or anticipated misalignments alongside proposed features, rather than single-function implementations.


\subsection{Difficulties in Resisting Technical Authority Across Multi-Level Value Tensions}

\subsubsection{Technical Authority and Extended Satisficing}

Using AI as reflective material produced a form of extended satisficing that differs from conventional \textit{good enough} decision-making in design. In traditional contexts, satisficing is often driven by time pressure or external constraints. In AI-mediated concept development, designers frequently accepted AI-generated concepts prematurely because suggestions about AI capabilities carry implicit technical authority. This authority emerged when AI outputs appeared both technically sophisticated and aligned with stated human values, a dynamic P3 captured, describing that “\textit{[the AI] gave me a list of ideas. It was through multiple back-and-forths, but the ideas seemed pretty solid and related to my chosen value, so I just went with them and started [working] off of them. But I realized later [that I had not] thoroughly explored the solution space},” leading to acceptance that felt justified in the moment, though later reflection revealed it as a false sense of completeness.

This pattern was not simply a matter of efficiency or lack of critical engagement. It reflected a distinctive epistemic situation in which one AI appeared qualified to define the design space for another AI. For instance, P3 articulated this realization, stating “\textit{when ChatGPT suggested machine learning as a behavior for my concept, it used some technical terms I recognized and could connect them to my values in such ways that seemed innovative. So, I took the suggestions without really thinking or knowing I was essentially letting one AI define the possibilities for another AI I was trying to design}.” As documented in our think-aloud notes, this pattern emerged across 13 of 18 participants during early ideation. Importantly, this authority was not entirely unwarranted. AI tools do possess genuine expertise about existing AI techniques and patterns, but that expertise is bounded by training data grounded in current systems and dominant paradigms. What made the authority particularly persuasive was its coupling of technical language with value framing, which obscured the limits of the suggestion space and masked potential harms or omissions.

Temporal pacing played a critical role in whether designers resisted or reinforced this authority. Participants who moved rapidly from value articulation to concept selection were more susceptible, while those who deliberately slowed down and revisited suggestions were more likely to surface gaps, risks, or misalignments. P14’s experience, described earlier, is a representative example. Resistance did not involve rejecting AI outright but rather interrupting the momentum through which authority solidified. This finding carries particular weight for less experienced designers. Generative AI systems tend toward outputs that affirm and extend the direction suggested by user prompts rather than introducing friction or challenge \cite{fu2024, liu2024}, and for designers with limited domain expertise or prior experience recognizing AI authority effects, this tendency may make extended satisficing nearly unavoidable without external intervention. The metacognitive demands of identifying and interrupting this dynamic are themselves substantial \cite{tankelevitch2024}, and reflective redirection, as an improvised strategy requiring awareness of the pattern before it can be disrupted, may be far less accessible to novices.

\subsubsection{Multi-Level Value Tensions and Harm Recognition}

A central finding was that designers using AI tools to design AI concepts had to manage values across three interdependent levels, namely values embedded in the AI tool, values guiding the designer’s intentions and intuitive harm sensibilities, and values intended for the designed concept. Unlike traditional design materials, AI systems encode assumptions, priorities, and normative commitments that actively shape ideation. These levels did not remain separate. They interacted continuously and created tensions and misalignments rather than stable alignment. Participants frequently struggled to locate the boundaries between these value sources, a confusion P13 captured well, noting “\textit{I wanted to design an AI that values human creativity, but I realized I was using ChatGPT, which has its own perspectives or programmed assumptions about creativity, to generate ideas about creativity-supporting AI. I couldn’t tell where my own values ended and ChatGPT’s began and vice versa}.” This reflected a distinctive epistemological challenge. When designers use an AI tool to reason about AI itself, the separation between tool values and designer values becomes porous and unstable. P1 articulated the vigilance required in this context, describing how designers “\textit{should be aware of toxic intent in AI and not blindly follow along. You must simultaneously be aware of values embedded in AI’s responses while also ensuring their own values are reflected in the AI system they’re designing}.” Decisions at one level constrain or enable possibilities at others, creating recursive loops rather than linear translation.

Across all activities, designers were consistently more attuned to value tensions, breakdowns, and potential harms than to articulating what successful value fulfillment should look like. Moments of discomfort, hesitation, or misalignment functioned as reliable signals that something was wrong. Designers could often identify when AI suggestions posed risks to agency, safety, or trust, even when they struggled to specify ideal alternatives (P1, P6, P9, P13). P9 described this dynamic when reflecting on AI-generated notions of user agency, noting a persistent feeling that “\textit{something was wrong with how the AI was suggesting I build user agency into the system, like it was creating the appearance of choice without real agency. I couldn’t articulate exactly why, but I knew it wasn’t right}.” This intuitive harm recognition proved more concrete and actionable than abstract value statements. These breakdowns were not treated as failures but as productive cues that redirected design.

Several participants began with clearly articulated values, then relied on AI for ideation, only to reject outputs that felt wrong, incomplete, or potentially harmful. This pattern suggests that designers’ intuition about what should not happen functions as a core design resource in AI-mediated practice. Across participants, values operated less as fixed principles to be implemented and more as evolving commitments clarified through recognizing misalignment. As P3 noted, “\textit{I can articulate how efficiency might manifest in interface design, like streamlined workflows, minimal friction, responsive feedback. But when I try to specify how it should guide the AI’s optimization strategies, I realize I need help to explore this further}.” In this way, harm recognition becomes a mechanism of knowledge production. In these cases, designers refined their understanding of values by resisting and reworking outputs that felt misaligned, though this was reactive and dependent on individual attunement rather than systematic practice, and should not be taken to imply that harm recognition as observed here reliably produces value-aligned concept outcomes.

When triangulating artifact analysis (see \hyperref[fig:concept-sketches]{Figure~\ref*{fig:concept-sketches}}) with interview data, we observed value tensions not surfaced during the activity or prominently reported in interviews. For example, P16’s sustainability-focused concept centers on an AI recommendation interface system that allows users to browse existing products and measure their sustainability metrics:
\begin{quote}
    “I imagine that it provides real-time detection of an existing product’s sustainability details, [suggests] how you can improve and make your products more sustainable, and predicts how sustainable a product will be. I want the interface to also include an AI agent in a chat box, like prompting what if I change from this material to a different material” (P16).
\end{quote} 
This envisioned concept would typically require continuous data collection and model retraining whose computational implications sit in potential tension with the stated sustainability value but go unaddressed in the sketch. While we do not claim that participants failed to recognize these tensions, the pattern is consistent with single-value framing enabling conceptual focus at the cost of suppressing cross-value conflicts that multi-value or adversarial framing might have surfaced earlier. Additional sketches suggest this pattern is not isolated. P8’s autonomy-centered concept depicts an AI-embedded design tool that mediates organizational design decisions by automatically detecting whether a certain element meets established standards. Yet the concept includes no visible mechanisms for supporting autonomy, such as override controls, intent-based operations, or adaptive feedback. P7’s efficiency concept presents a similar case of AI-automated workflow transfer across devices with no visible human checkpoints. Taken together, these artifact observations suggest the activity’s single-value structure enabled focus at the cost of rendering cross-value tensions invisible during ideation itself.

These dynamics also shaped how designers reasoned about agency. Choices about how much authority to grant the AI tool during ideation became entangled with decisions about how much agency the designed system should later exercise over users. For instance, P6 reflected, “\textit{I used AI to tell me what problems that designers on a whole are experiencing. Granted, I don’t know where AI got that information from. I guess I didn’t really question it. If I let ChatGPT make too many decisions about my AI concept, am I unconsciously designing an AI system that will make too many decisions for its users then?}” (P6). Their concern emphasizes that values embedded in the ideation tool risk propagating into the designed concept’s form and features, user flows, and interaction patterns. For designers, the design process itself functions as a prototype for the human-AI relationship they were embedding.

Value interference was not uniformly negative. In some cases, it forced confrontation with assumptions that might otherwise remain invisible. P16 captured this tension, describing persistent uncertainty about whether the AI tool’s implicit values about recommendation systems were subtly reshaping sustainability goals, asking “\textit{whether the AI I’m using had implicit values about recommendation algorithms that were influencing my design choices in ways that might conflict with my intended values for my new concept, given that I chose to design an AI recommendation interface centered on sustainability}.” While cognitively demanding, these moments sometimes enabled deeper reflection and earlier harm anticipation. Participants also developed nuanced judgments about when AI authority should remain secondary. P11 stated that “\textit{AI might be appropriate for standard design approaches that do not require a lot of creativity. But contexts like museums or politics require human-centered vision that brings surprises beyond current generative AI capabilities}.” These judgments are not abstract principles but situational assessments grounded in perceived stakes, including potential for harm, loss of agency, or erosion of trust.


\subsection{Challenges and Emerging Strategies in Interpretation-Centered Design Practice}

\subsubsection{Interpretation-Centered Design Expertise}

Engagement with AI as reflective material precipitated a shift from execution-focused toward interpretation-centered design practice. Participants increasingly spent their effort assessing possibilities, weighing value trade-offs, and making decisions about concept direction, rather than thinking about interface-based elements and sketching perfect product features. As P12 described, the work involved “\textit{moving less like pixel pushing and doing, practicing, and thinking more about the value, impact, and strategy}.” Design expertise is increasingly exercised through judgment rather than execution.

Creativity emerged as a distinctively human form of expertise within this interpretive work. Participants did not describe creativity as visual novelty but as the ability to intentionally depart from rules and expectations. P9 characterized creativity as “\textit{human intervention of purposefully breaking a rule},” a capacity participants felt was difficult to delegate to AI. Several designers described monitoring whether AI outputs reinforced familiar patterns or converged too quickly on predictable solutions, and intervened when distinctiveness was lost. P16 emphasized caring “\textit{more about making the solution unique}” because “\textit{everyone can eventually think of and reach [AI-generated ideas]}.”

Participants also distinguished between AI-generated variation and creative direction. While randomness is sometimes useful, it requires human interpretation to become meaningful. P8 captured this directly, noting that “\textit{true randomness would seem like garbage to us},” and that its value came from how designers interpreted and redirected it. Designers treated AI variation as material to work with rather than as creative output in itself, selectively amplifying or suppressing it based on whether it supported their intended values, consistent with work demonstrating that semantically distant or random ideation inputs require human interpretive work to become generatively useful \cite{chan2017semantically}, and with accounts of the distinction between experiential and reflective thinking in making sense of tool-generated outputs \cite{norman2014}.

Participants with strong visual design backgrounds initially experienced this transition as a professional identity challenge. P5 reflected, “\textit{I’ve spent years developing visual design skills, like typography, layout, color theory. When AI can generate those elements, what’s left of my expertise? I felt like my professional identity was being questioned}.” Several others, however, saw this differently. As P14 explained, “\textit{I realized my visual design skills weren’t becoming irrelevant. They were becoming evaluation and curation skills. I can judge AI-generated visuals more sophisticatedly than non-designers because I understand the principles behind good design, shifting from making to critiquing and refining}.” Maintaining interpretive authority required designers to resist the tendency to accept AI outputs at face value, particularly when outputs appeared technically polished or persuasive. This reframing is intuitive and well-articulated, but it represents participants’ retrospective accounts of an emerging shift in professional identity rather than demonstrated evidence of sustained interpretive effectiveness in practice.

\subsubsection{Meta-Design Awareness}

As design expertise shifted toward interpretation, some participants found themselves managing substantial metacognitive demands, simultaneously reasoning about the AI tool, the design process, and the AI-enabled system they were envisioning, a complexity already documented in AI-mediated ideation contexts \cite{anderson2024, fu2024, sinlapanuntakul2025SLR, tankelevitch2024}. This awareness is distinctive to designing AI using AI, where tool assumptions and limitations risk unconsciously transferring into designed concepts. P10 articulated this clearly, noting “\textit{Designing AI using AI requires meta-level understanding, which isn’t necessary when using AI for other non-AI design challenges}.” Meta-design awareness manifested through strategic judgment about when to accept, modify, or reject AI suggestions. As P6 noted, 
\begin{quote}
    “Using AI isn’t all about prompt engineering but more about managing a strategic conversation. I need to have the instincts for when to push back on its suggestions, when to redirect the conversation, and when to take its multiple outputs and combine them into something completely new. These are judgment calls that require design expertise, not technical skills” (P6).
\end{quote}
This judgment extends beyond prompt formulation to maintaining value alignment across multiple levels, including designers’ guiding values, perceived values embedded in AI tools, and intended values of the concept.

Notably, participants were more consistently attuned to recognizing breakdowns, misalignments, and emerging harms than to articulating abstract positive value fulfillment. P4 described moments when AI authority threatened to override independent judgment, noting that AI suggestions can “\textit{sound so authoritative that they can override my design judgment. I had to learn to remind myself when I was being totally persuaded by AI authority rather than making independent strategic decisions}.” These experiences of discomfort functioned as concrete signals that value alignment had shifted or failed. Rather than treating such breakdowns as errors to be corrected later, designers used them to anticipate potential harms and redirect concepts early in development.

Participants also described how meta-design awareness supported more speculative, future-oriented reasoning. When envisioning AI-enabled concepts, designers were not only resolving present value tensions but also anticipating how those values might shift as systems are deployed and adapted over time. P10 used a historical analogy to make sense of this uncertainty:
\begin{quote}
“When the camera was invented, a lot of the painters were really mad because they stopped getting commissions for portraits. They used to be the ones making realistic pictures, but once the camera came into play, everybody could just have their own pictures. That kind of decentralized production and image making, but then that just created its own kind of creativity, which is why we have photography as a medium today” (P10).
\end{quote}
For P10, AI similarly reconfigures what counts as design work and expertise. Rather than eliminating creativity, it shifts where and how values are negotiated, opening up new possibilities that only become visible through forward-looking speculation. Several participants described having to imagine how their own value commitments might need to change as concepts move from speculation to use. As P8 put it, “\textit{When designing AI systems for innovative contexts, I’m speculating about how my values might need to evolve or adapt. The AI tool can help me think through current value conflicts, but I have to imagine how those conflicts might change as my concept is implemented and adopted}.” In this process, asking one AI to imagine another exposed embedded assumptions in current AI paradigms, surfacing risks and constraints that would remain invisible in non-AI design tasks.


\section{Discussion}

Our findings reveal that designing \textit{with} generative AI \textit{for} AI concepts reshapes reflection-in-action, introduces new forms of technical authority, and reframes how values are articulated and negotiated in design, with implications for design discussed in this section.


\subsection{Extending Reflection-in-Action}

Schön’s concept of reflection-in-action frames design as a conversation between a designer and their materials, where actions produce consequences that guide subsequent moves \cite{schon1983, schon1987}. In this account, designers engage in a \textit{conversation with the materials of a situation} \cite{schon1992}. Materials talk back through resistance, affordance, and surprise, prompting continuous reframing and adjustment in a process that is already bidirectional and responsive. However, the conversation remains asymmetric; meaning-making is primarily human-driven, and materials reveal possibilities through physical properties and structural constraints rather than through semantic interpretation. Clay does not infer the sculptor’s intent, generate alternative formal proposals, or embed normative assumptions about what the work should become. Interpretation belongs to the designer; the material responds, constrains, and discloses \cite{doordan2003, schon1983, schon1992}. It is precisely this asymmetry that separates conventional material engagement from AI, where the material returns not physical resistance but semantic interpretation and generative agency \cite{holmquist2017, benjamin2021, lindgren2023}.

Our findings show that this assumption of interpretive asymmetry does not hold when designers work with generative AI. In early-stage AI concept design, the material does not merely respond but actively interprets, reframes, and proposes. AI systems infer designer intent, translate values into technical possibilities, and suggest directions for further exploration. As a result, reflection no longer unfolds as a unidirectional conversation guided by material feedback. Instead, it becomes a recursive dialogue in which both designer and tool participate in interpretation and generation. We refer to this process as reciprocal reflection-in-action, a qualitatively different structure of reflection that extends Schön’s framework to account for materials with interpretive agency.

This difference manifests in how designers engage with generative AI’s outputs. Participants did not treat generated suggestions as neutral responses but as interpretations requiring evaluation. Reflecting on this process, participants noted that watching how an AI interpreted \textit{efficiency} as a value and translated it into product features, and then judging whether those features reflected what \textit{efficiency} meant to them, required reflection at multiple levels. Through this process, designers encountered their own values indirectly, mediated by the tool’s reasoning, rather than through direct manipulation of form or structure. This layered sensemaking contrasts with traditional reflective practices such as sketching or note-taking, where external representations primarily function as extensions of the designer’s own thinking \cite{goldschmidt1991, purcellandgero1998, doordan2003}.

Prior research has emphasized that design is exploratory and that goals often emerge through the act of designing itself, particularly when designers work with ill-defined problems \cite{buchanan1992}. Our findings extend this understanding by showing what changes when the material itself participates in sensemaking. Interpretive materials expand the space of possible reframings by proposing directions designers may not have considered. At the same time, they introduce new cognitive demands. Designers must evaluate not only the novelty or feasibility of ideas but also the assumptions embedded in how the tool frames problems, values, and capabilities. Schön described designers as developing a tacit feel for how materials behave under intervention \cite{schon1983}. With generative AI, designers develop sensitivity to how the material interprets intent, what defaults shape its outputs, and where those defaults systematically narrow imagination, suggesting a form of reflection Schön’s original framework did not anticipate.

This framework also clarifies how designer agency is exercised in AI-mediated design. Prior work has shown that effective AI co-creation depends on designers maintaining judgment and direction while working with generative systems \cite{guo2023, hu2025, sinlapanuntakul2025SLR}. Our findings extend this work by identifying specific mechanisms that challenge and enable agency. Designers encountered technical authority, value interference, and recursive conventional thinking that threatened to override independent judgment. In response, they developed strategies such as slowing down interaction, creating distance from initial outputs, and using moments of discomfort as signals for further reflection. These findings align with recent work on AI-supported ideation that highlights both the generative benefits and the cognitive overwhelm introduced by accelerated idea production, offering concrete mechanisms through which designers can maintain critical perspective \cite{fu2024, liu2024, palani2024, rayan2024, sinlapanuntakul2025SLR}.

The recursive condition of using AI to design AI introduces particular complexity. Designers must reason simultaneously about their own agency relative to the tool, the tool’s agency in generating concepts, and the agency they intend to embed in the systems they are designing. This form of nested agency introduces cognitive complexity that existing design cognition theories do not fully account for \cite{anderson2024, fu2024} and represents a novel challenge distinct from using AI for non-AI design tasks \cite{mccormack2020, shi2023, palani2024, hu2025, oppenlaender2025, sinlapanuntakul2025SLR}. Recent work on LLM-supported design space exploration \cite{zamfirescu2025} demonstrates how generative AI shapes which possibilities feel tractable during concept development. Our findings extend this by showing that shaping operates through value-laden framing rather than capability disclosure alone. Work examining how tools configure power relations in design practice \cite{li2024} resonates with our finding that AI’s technical authority is difficult to resist precisely because it arrives coupled with apparent value alignment, making its influence on concept direction feel credible and self-evident rather than imposed.


\subsection{Technical Authority vs. Traditional Satisficing}

Prior work on design decision-making has documented satisficing, defined as the practice of accepting solutions that appear adequate rather than exhaustively searching for optimal ones \cite{simon1996, cross2021}. This behavior is commonly explained through constraints such as limited time, cognitive load, or resource scarcity. Our findings point to a different mechanism operating in AI-mediated design, one that cannot be fully explained by these traditional accounts.

We observed a form of deference rooted in perceived technical authority rather than convenience. This authority emerged from the convergence of two factors. First, the AI signaled expertise in AI systems through technical language and familiar references that designers recognized as legitimate. Second, its suggestions often appeared to align with the values designers had already articulated, reinforcing the perception that the AI’s recommendations were credible. Here, acceptance did not stem from time pressure but from the perceived coherence between technical plausibility and value alignment.

Crucially, the cues that led participants to trust the AI did not reliably indicate its actual trustworthiness in this context. Although AI training on existing systems supports fluency with current techniques, it also limits the range of possibilities the system can plausibly generate \cite{sinlapanuntakul2025SLR, fu2024}. The same data that enables technical fluency simultaneously reproduces dominant design conventions. Thirteen of the eighteen participants encountered this tension, suggesting that misplaced trust in technical authority is not an exceptional case but a widespread, consequential challenge in AI-mediated design.


\subsection{Reframing Values in AI Development}

Our findings complicate prevailing accounts of how values operate in responsible AI development \cite{holstein2019, umbrello2021}. Rather than functioning as fixed constraints or ethical checklists, values emerged as evolving commitments shaped through iterative engagement with AI tools \cite{friedman2019} and through moments of discomfort and misalignment. Designers worked with values as hierarchical, situated systems grounded in lived professional experience \cite{cross2021, friedman2002}, producing secondary manifestations specific to the meta-design context of using AI to design AI. This contrasts with the abstract and universal treatment of values common in existing frameworks \cite{sadek2024designing, sadek2024guidelines} and underscores the relational and experiential character of effective value integration \cite{benjamin2021, dove2020, tian2024}.

Current AI development practices frequently position ethics as a late-stage intervention or post-hoc auditing activity \cite{dove2017, yildirim2024}, addressing values only after system architectures have solidified or harms have surfaced \cite{raji2020}. Such reactive approaches limit the extent to which values can meaningfully shape design decisions \cite{friedman2019, gabriel2020, umbrello2021}. Our findings instead point toward a shift in orientation, where values operate as proactive design materials that actively inform system conceptualization from the outset. This requires iterative and exploratory engagement, in which values are negotiated through use rather than implemented as fixed principles. While aligned with recent calls for values-centered AI design \cite{gabriel2020, holstein2019, shen2024bidirectional, shen2024valuecompass, stray2024}, our study highlights how these challenges intensify when designers work directly with AI systems during concept development.

Our findings also foreground the temporal complexity of value integration in AI systems, a dimension underexplored in current design frameworks. Unlike conventional interactive systems with relatively stable interaction patterns, AI systems evolve through accumulated user interactions, contextual feedback, and learning over time \cite{son2024}. As a result, values embedded in AI cannot be understood solely through immediate interactions. Designers must also account for how values may shift as systems adapt to new users, contexts, and edge cases. For example, an AI-enabled product initially designed around ownership might first express this value through customization and user control, yet later reinterpret ownership as transparency or revision and correction affordances based on emergent use patterns. Participants consistently struggled to anticipate these forward-looking value transformations, revealing a gap in prevailing design practices that focus on discrete interaction moments rather than evolving value relationships.

Critically, our findings suggest that harm recognition functions as a concrete and actionable resource within this value negotiation process. Rather than relying on abstract articulation of positive value ideals, which designers found difficult, they engaged substantively through noticing misalignment, recognizing discomfort, and anticipating breakdowns \cite{namer2025harm96designers}. Sensitivity to what should not happen offered designers a grounded mode of ethical reasoning, one that supported continuous adjustment as AI concepts evolved \cite{gabriel2020, holstein2019, stray2024}. In this sense, ethical engagement is not separate from design work but embedded within it, unfolding through iterative interaction with AI systems. Our findings also connect with work on value tensions in design practice. Sabie et al. \cite{sabie2022}’s examination of unmaking as a design stance and Haghighi et al. \cite{haghighi2023}’s workshop-based methods for navigating value tensions both suggest that productive engagement with values in design requires deliberate space for contradiction, resistance, and redirection, which is consistent with our finding that harm recognition and reflective redirection were more reliable design resources than positive value articulation.

Across these dynamics, values did not guide design in a linear or harmonious manner. When brought into contact with concrete design materials, technical constraints, and AI-mediated interpretations, values generated tensions and conflicts visible only through engagement \cite{friedman2019, shen2024bidirectional, shen2024valuecompass}. These moments of friction were particularly pronounced when AI tools actively interpreted, translated, or reframed value concepts, often surfacing misalignments through use rather than deliberation. Rather than treating such breakdowns as errors to be corrected after the fact, our findings suggest that harm recognition plays a continuous, central role in ongoing value negotiation. This reframes ethical work from a phase-based activity to a constitutive feature of how designers think through AI systems from inception.


\subsection{Design Implications}

Our findings suggest several critical implications for advancing design practices that support value-oriented AI concept development.

\paragraph{Rethinking AI design tools} Existing AI design tools should be fundamentally reimagined to scaffold reflection-in-action and multi-level value tracking rather than optimizing solely for productivity and frictionless interaction. Prior work shows that prompt–response interaction paradigms limit designers’ ability to engage in abductive loops of interpreting, reframing, and re-grounding ideas during concept development \cite{rezwana2023}. While current tools maximize output quality and minimize user effort, making AI interaction feel frictionless, our findings suggest that friction, in the form of structured reflection pauses \cite{yildirim2022, yildirim2023pairguidebook, anderson2024}, can support deeper reasoning about breakdowns, misalignment, and value conflict rather than smoothing them over. Designers were particularly attuned to moments when something felt wrong \cite{namer2025harm96designers}. Tools should create explicit opportunities to capture and explore these moments, helping designers pause when they sense something is wrong and use such moments as signals for deeper thinking \cite{liao2023}.

\paragraph{Complementary non-AI scaffolding} Our findings reveal a critical need for non-AI-powered tools specifically designed to scaffold AI concept development. While AI tools are valuable for generative exploration, their interpretive authority and generative fluency can overwhelm reflective processes, particularly when designers are working on innovative or value-laden concepts \cite{churchill2025, raisamo2019}. We observed that designers who slowed down, created distance from initial AI responses, and deliberately imagined alternatives engaged in more critical and creative reasoning. Crucially, these toolkits should center values respected and promoted by design practice \cite{umbrello2021}, providing dedicated scaffolding for value articulation throughout the design process. Non-AI tools provide stable cognitive anchors and create space for critical distance from AI-generated outputs. In the absence of such tools, reflection risks becoming superficial and driven by the persuasive authority of AI outputs rather than deliberate judgment. However, our findings on harm recognition suggest that designers often engaged substantively through noticing misalignment, discomfort, or potential harm rather than through formal value frameworks. Accordingly, complementary non-AI frameworks should explicitly support this form of negative reasoning by helping designers articulate what should not happen, recognize emergent harms, and anticipate breakdowns. Rather than replacing AI tools, these approaches should be understood as complementary, with AI supporting generation and exploration, and non-AI scaffolding supporting structured reflection, deliberation, and harm anticipation. This complementary framing is one we operationalized in developing our AI Concept Envisioning toolkit \cite{sinlapanuntakul2026toolkit}.

\paragraph{Organizational positioning of design} Our findings extend beyond design practice to argue that design must be integrated into human-centered AI development from inception, requiring fundamental organizational reconsideration. Rather than positioning design primarily as a user experience layer applied after system architecture decisions have been made, organizations should position design as foundational to AI development \cite{benjamin2021, feng2023, verganti2020, yang2018UXML, yang2020, zdanowska2022, zimmerman2020}. This requires giving designers genuine influence over what problems get solved and what solutions get attempted, not just how solutions are executed. It requires creating structures where designers can engage with engineers, ethicists, and domain experts in genuine collaboration \cite{nahar2022}. This is organizationally challenging but conceptually important. It acknowledges that how designers approach value articulation, concept alignment, and harm recognition during early ideation directly shapes what becomes possible or constrained in later development phases. Our findings show that designers bring distinctive capabilities in recognizing when something might be wrong or harmful. Positioning design as foundational means creating organizational structures that value and amplify these capabilities rather than treating design as a later-stage refinement layer.


\section{Limitations and Future Research} 
\label{limitations}

Our study has limitations that should be considered when interpreting the findings. The 20-minute design activity captures early-stage ideation in a specific context and cannot capture the full process of AI concept development. The limited session time reduced opportunities to revisit and renegotiate value commitments as concepts developed, precisely when cross-value tensions tend to surface. Asking participants to engage with only a single value, while ecologically valid, further bounded the kind of deep value tension exploration that more extended or multi-value engagements would enable. These choices reflect our scoping toward naturalistic early-stage ideation rather than structured value negotiation, and findings should be interpreted accordingly. Nevertheless, understanding how designers approach value integration and harm recognition at this formative stage matters precisely because it establishes foundational direction that constrains or enables possibilities in later development phases \cite{yildirim2024sketching}. Future work could build on this by examining how extended multi-session engagement and iterative reflection influence design outcomes.

We also allowed participants to use generative AI tools of their choice to preserve ecological validity and reflect real-world variation in workflows. As a result, tool-specific factors such as interaction history and model version likely shaped the character and quality of AI outputs participants received. Future work should standardize the AI tools and account for user familiarity levels to isolate tool-specific influences.

While all participants reported prior interaction with AI tools, their experiences ranged from end-user engagement to active involvement in AI product design, and we did not assess their prior orientations toward AI. Familiarity with AI does not necessarily translate to understanding its inner workings \cite{yang2018UXML}, and those who had developed more critical stances through professional experience may have been more attuned to recognizing AI authority effects than those with more uniformly positive associations. Participants’ varying familiarity with explicit value reflection also likely shaped the depth of engagement visible in our analysis. Future work should examine these dynamics across varying levels of AI familiarity and prior orientation, and could explore how targeted resources might deepen designers’ comprehension of AI’s capabilities and constraints.

Finally, while this work focused on design practitioners, AI product concept development in practice often involves multidisciplinary teams including product managers, data scientists, and others. Building on recent research \cite{deng2023, kross2021, mao2019, muralikumar2025, nahar2022, yildirim2023pairguidebook, zhang2020}, future work could explore role-specific support mechanisms and resources, moving beyond general collaboration facilitation toward tailored interventions that address each role’s distinct contributions and needs.


\section{Conclusion}

This study explored how designers navigate the dual role of using AI tools to envision AI-enabled concepts, revealing how doing so generates recursive value tensions that differ fundamentally from traditional design practice. Through a design activity and reflective interviews with 18 designers, we identify reciprocal reflection-in-action comprising three interconnected modes of engagement, namely value externalization, interpretive encounter, and reflective redirection, extending Schön’s work to account for design materials with interpretive agency. Our findings further show that designers are more reliably attuned to recognizing potential harmful effects and value misalignments than to articulating abstract notions of positive value fulfillment. This harm-recognition capability functions as a concrete and actionable design signal, guiding reflective redirection and sustaining human interpretive authority in contexts that pressure designers to defer to AI-generated suggestions. Our study suggests that responsible AI design requires tools, methods, and organizational structures that support ongoing value negotiation in early-stage concept design rather than post-hoc auditing, positioning design as foundational to how AI-enabled products are conceived and developed over time.


\begin{acks}

This research was supported by a doctoral research grant from Ramey Research Fund at the University of Washington’s Department of Human Centered Design \& Engineering.

\end{acks}


\bibliographystyle{ACM-Reference-Format}
\bibliography{references}


\onecolumn
\appendix

\section{Codebook} 
\label{appendix:codebook}
\input{materials/codebook}

\section{Analysis of Relevant Parties} 
\label{appendix:relevant-parties-analysis}
\begin{landscape}
\input{materials/relevant-parties-analysis}
\end{landscape}


\end{document}

%% file: materials/participants.tex
\begin{table*}[!t]
    \caption{Participants’ demographic information.}
    \Description{Table displaying self-reported demographic data of 18 participants, including their professional role, years of experience, design expertise, and educational background.}
    \label{tab:participants}
    \centering
    \small
    \renewcommand{\arraystretch}{1.2}
    \begin{tabular}{lllll}
    \toprule
        \textbf{ID} & 
        \textbf{ Professional Role} & 
        \textbf{Exp. (yrs)} & 
        \textbf{Design Expertise} & 
        \textbf{Educational Background} \\
    \midrule
        P1 & Experience designer & 2 & AR/VR & Human-computer interaction \\
        P2 & Experience designer & 5 & B2C & Human-computer interaction \\
        P3 & UX designer & 2.5 & Experience design & Interaction design \\
        P4 & Senior UX designer & 3.5 & B2B & Human-computer interaction \\
        P5 & Senior UX designer & 12 & AI and education & Visual communication \\
        P6 & Product design consultant & 3 & Early-stage startups & Political science \\
        P7 & Interaction designer & 2 & Speculative design & Human-computer interaction \\
        P8 & Staff product manager & 24 & Organization design & Computer engineering \\
        P9 & Senior product designer & 10 & Product design & Human-computer interaction\\
        P10 & Interaction designer & 3 & UI design & Design and technology\\
        P11 & Design researcher & 6 & Human-robot interaction & Industrial design \\
        P12 & Lead UX designer & 3 & Data visualization & Kinesiology \\
        P13 & Design lead & 7 & AR/VR & Human-computer interaction \\
        P14 & Product design manager & 20 & AI & Visual communication \\
        P15 & CEO / Design director & 21 & Child-centered design & Interaction design \\
        P16 & Product designer & 4 & Computer networking & Human-computer interaction \\
        P17 & UX designer & 2.5 & E-commerce enterprise & Human-computer interaction \\
        P18 & Experience designer & 3 & Accessibility & Human-computer interaction \\
    \bottomrule
    \end{tabular}
\end{table*}

%% file: materials/concept-sketches.tex
\begin{figure*}[!ht]
    \centering
    \includegraphics[width=\linewidth]{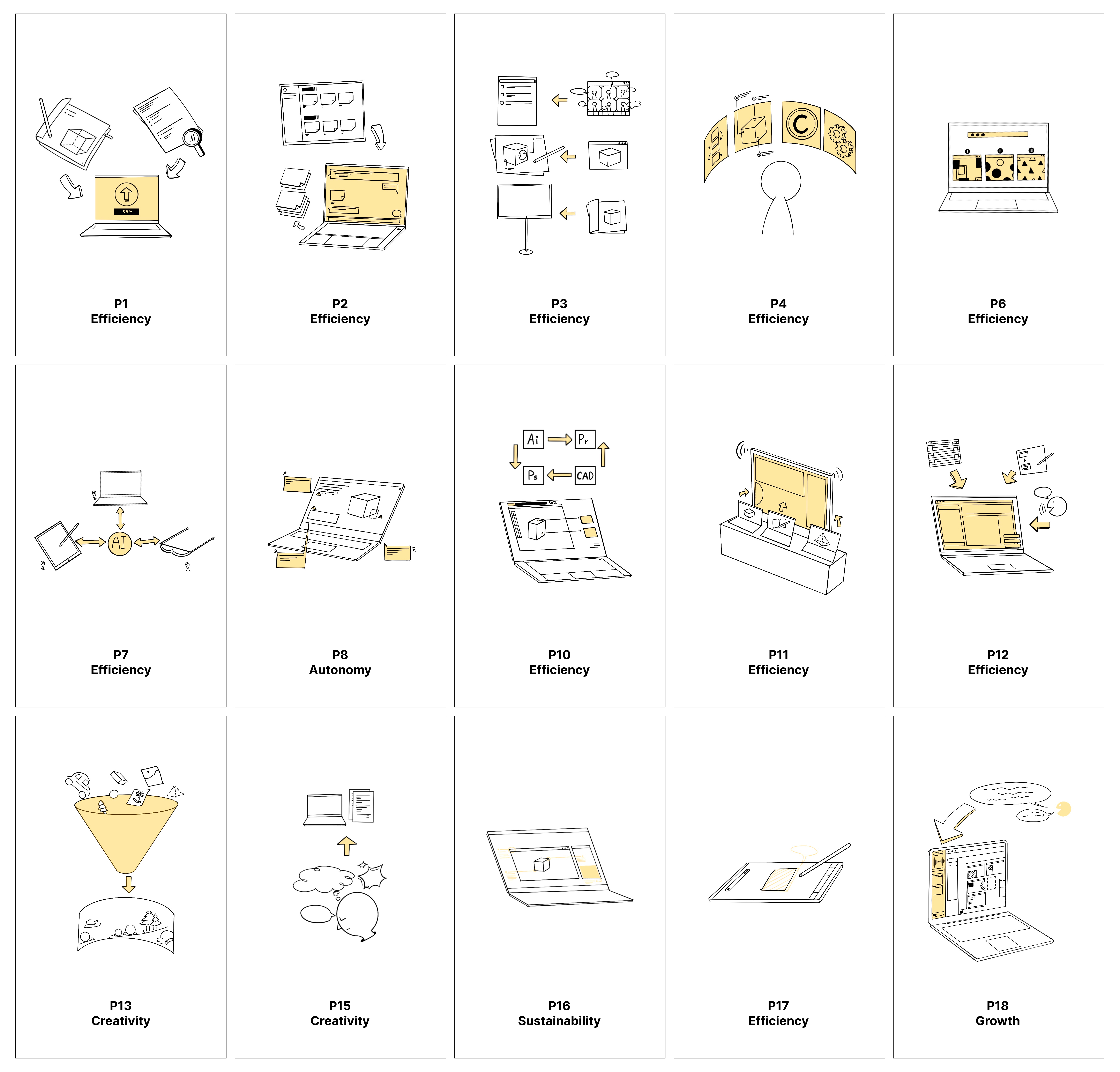}
    \caption{Examples of AI concept sketches. We re-sketched 15 of 18 participant sketches for clarity and excluded 3 sketches whose visual and textual descriptions were too ambiguous for reliable interpretation. Sketches represent participants’ final concept representations (at the end of the activity), annotated with the value each participant selected at the start. Although values are shown as fixed labels, participants’ interpretations of their selected values evolved throughout the design process.}
    \Description{Examples of AI concept sketches. 15 of 18 participant sketches were re-illustrated for clarity, and 3 sketches whose visual and textual descriptions were too ambiguous for reliable interpretation were excluded. Sketches represent participants’ final concept representations (at the end of the activity), annotated with the value each participant selected at the start. Although values are shown as fixed labels, participants’ interpretations of their selected values evolved throughout the design process.}
    \label{fig:concept-sketches}
\end{figure*}

%% file: materials/RRiA.tex
\begin{figure}[H]
    \centering
    \includegraphics[width=\linewidth]{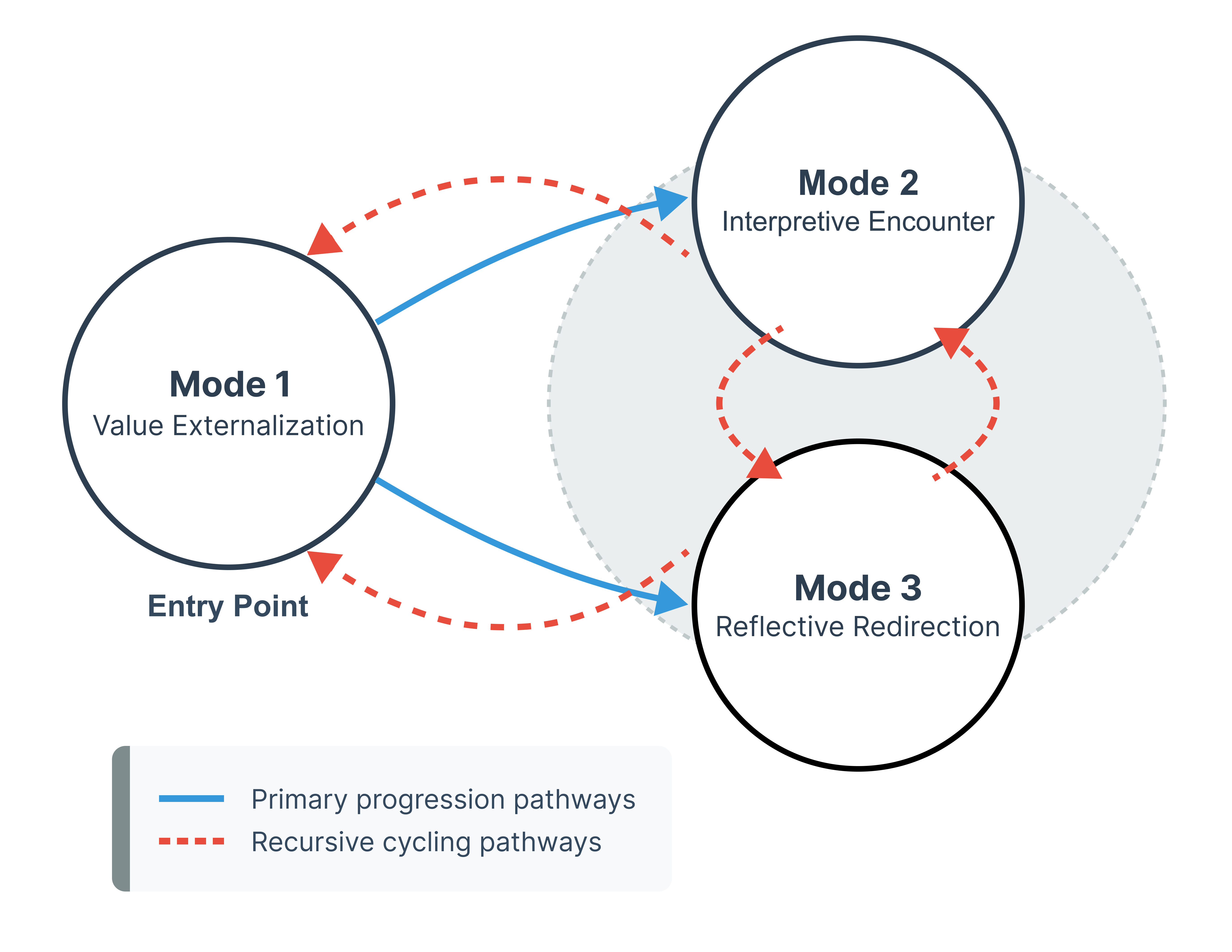}
    \caption{Process Framework of Reciprocal Reflection-in-Action with AI, which depicts three non-sequential, mutually reinforcing modes of engagement observed when designers used generative AI for AI-enabled concept development: value externalization (mode 1), interpretive encounter (mode 2), and reflective redirection (mode 3) with primary progression pathways from mode 1 to modes 2 and 3 as well as recursive cycling pathways between all the modes.}
    \Description{Process Framework of Reciprocal Reflection-in-Action with AI, which depicts three non-sequential, mutually reinforcing modes of engagement observed when designers used generative AI for AI-enabled concept development: value externalization (mode 1), interpretive encounter (mode 2), and reflective redirection (mode 3) with primary progression pathways from mode 1 to modes 2 and 3 as well as recursive cycling pathways between all the modes.}
    \label{fig:RRiA}
\end{figure}

%% file: materials/codebook.tex
\begin{table}[!h]
    \captionof{table}{Example of the final codebook, including codes, subcodes, and their given definitions.}
    \Description{Example of the final codebook used in the analysis, including codes, subcodes, and their defined definitions.}
    \label{tab:codebook}
    \centering
    \small 
    \renewcommand{\arraystretch}{1.15}
    \begin{tabular}{lll}
    \toprule
        \textbf{Code} & 
        \textbf{Subcode} & 
        \textbf{Definition} \\
    \midrule
        \multirow{3}{*}{AI in Design}
        & Opportunities 
        & New design possibilities enabled by the advancement of AI. \\ [0.15cm]
        & Concerns 
        & Considerations to navigate when using AI in design work. \\ [0.15cm]
        & Limitations 
        & Constraints related to the capabilities and issues of AI tools. \\ [0.15cm]
    \midrule
        \multirow{2}{*}{Designing AI}
        & Ideation 
        & Action-based factors when ideating solutions. \\ [0.15cm]
        & Selection 
        & Action-based factors when selecting a solution. \\ [0.15cm]
    \midrule
        \multirow{6}{*}{Role of AI}
        & \multirow{2}{*}{Partner/collaborator}
        & The perception of AI as an equal participant in the process, \\ & & contributing ideas and solutions alongside human designers. \\ [0.15cm]
        & \multirow{2}{*}{Tool}
        & AI as a supportive tool that aids designers in their tasks \\ & & and workflows without taking over creative responsibilities. \\ [0.15cm]
        & \multirow{2}{*}{Idea management}
        & The process by which designers expand their thinking and \\ & & reflect on their practices through interaction with AI tools. \\ [0.15cm]
    \midrule
        \multirow{8}{*}{Value}
        & \multirow{2}{*}{Manifestation}
        & Ways in which designers (un)intentionally insert \\ & & values into their design thinking and outcomes. \\ [0.15cm]
        & \multirow{2}{*}{Evolution}
        & How the experience of using generative AI in design \\ & & shapes or transforms designers' values and priorities. \\ [0.15cm]
        & Tension
        & The balancing act between competing values in design. \\ [0.15cm]
        & \multirow{2}{*}{Agency}
        & The extent to which designers feel in control of the design \\ & & process and the ownership they attribute to their creations. \\ [0.15cm]
    \bottomrule
    \end{tabular}
\end{table}

%% file: materials/relevant-parties-analysis.tex
\begin{table}[h]
\caption{Analysis of relevant parties, including direct and indirect parties along with their associated value(s) and value tension(s).}
\Description{Table displaying relevant parties analysis with potential direct and indirect parties, values, working definitions, benefits, harms, and value tensions.}
\label{tab:relevant-parties-analysis}
\centering
\small
\renewcommand\tabularxcolumn[1]{m{#1}}
\begin{tabularx}{\linewidth}{
    >{\raggedright\arraybackslash}m{1.75cm}
    >{\raggedright\arraybackslash}m{1.75cm}
    >{\raggedright\arraybackslash}m{3.75cm}
    >{\raggedright\arraybackslash}X
    >{\raggedright\arraybackslash}X
    >{\raggedright\arraybackslash}X
}
\toprule
\textbf{Parties} & \textbf{Value} & \textbf{Working definition} & \textbf{Benefit} & \textbf{Harm} & \textbf{Value tension} \\
\midrule

Designers (direct) &
Adaptability &
Ability to stay up-to-date with the latest tools in the field &
Working with constantly developing AI-enabled tools can help designers adapt faster &
Rapid changes in AI development could be overwhelming and cause stress &
The need to adapt to new AI systems/tools vs. potential stress and learning curve \\ [0.6cm]

&
Agency &
Ability of designers to make independent decisions during the design process &
Data-driven insights that inform the designers' design decisions &
AI starting to dictate the design process &
AI informing decisions vs. preserving the designers' agency in the design process \\ [0.6cm]

&
Creativity / Innovation &
Ability to come up with original and innovative ideas &
New perspectives and ideas that designers may have missed &
Over-reliance on AI could hold a designer's own creativity and innovative thinking back &
AI enhancing creativity vs. maintaining the designer's unique creative voice \\ [0.6cm]

&
Efficiency &
Reducing time and effort spent on design tasks while maintaining quality &
Faster completion of design projects and iterations and increased output without sacrificing quality through AI automation of repetitive tasks &
Neglecting crucial design considerations or overlooking unexpected challenges &
Maximizing output vs. ensuring the design outcome addresses all relevant aspects and aligns with ethical considerations \\ [0.85cm]

&
Growth &
Personal and professional development &
Working with AI can provide new learning opportunities and skills &
Too much automation from AI in the design process limits the designers' room for growth &
Ensuring the designers continue to grow and learn while integrating AI into their design workflow \\ [0.6cm]

\midrule

The Company (direct) &
Profitability &
Aim to create a product that is successful in the market &
Increased efficiency and innovation lead to a more competitive product &
Financial losses from misalignment between AI outputs and market needs &
Pursuit of innovative AI technology vs. market demands and profitability \\ [0.6cm]

&
Reputation &
Standing and credibility in the market/society &
Successful integration of AI in design leads to reputation as one of the technology leaders &
Ethical missteps or public failures with AI harm the company's reputation &
Managing the risk and reward of being an early adopter of AI in design \\ [0.6cm]

\midrule

AI Developers (indirect) &
Technological advancement &
Pushing the boundaries of what is possible with AI &
Developing AI for design tools can lead to new advancements for the society &
Ethical concerns and unintended consequences &
The drive for advancement vs. the need for careful, ethical development practices \\ [0.6cm]

\midrule

AI-Designed Product Users (indirect) &
Functionality &
The product should serve the users' needs effectively &
AI can potentially lead to better, more user-centric designs based on its pattern-recognition ability in user experience and behavior &
AI not fully understanding user needs could lead to ineffective products &
Ensuring the product functionality aligns with user needs while leveraging AI capabilities \\ [0.85cm]

&
Trust &
Users need to trust the product and the company behind it &
Transparent use of AI in design can build user trust &
Misuse of AI or lack of transparency erode trust &
The use of AI in design vs. the need to build and maintain user trust \\ [0.6cm]

\bottomrule
\end{tabularx}
\end{table}